\overfullrule = 0pt

\font\bigg=cmbx10 at 17.3 truept    \font\bgg=cmbx10 at 12 truept
\font\twelverm=cmr10 scaled 1200    \font\twelvei=cmmi10 scaled 1200
\font\twelvesy=cmsy10 scaled 1200   \font\twelveex=cmex10 scaled 1200
\font\twelvebf=cmbx10 scaled 1200   \font\twelvesl=cmsl10 scaled 1200
\font\twelvett=cmtt10 scaled 1200   \font\twelveit=cmti10 scaled 1200
\def\twelvepoint{\normalbaselineskip=12.4pt
  \abovedisplayskip 12.4pt plus 3pt minus 9pt
  \belowdisplayskip 12.4pt plus 3pt minus 9pt
  \abovedisplayshortskip 0pt plus 3pt
  \belowdisplayshortskip 7.2pt plus 3pt minus 4pt
  \smallskipamount=3.6pt plus1.2pt minus1.2pt
  \medskipamount=7.2pt plus2.4pt minus2.4pt
  \bigskipamount=14.4pt plus4.8pt minus4.8pt
  \def\rm{\fam0\twelverm}          \def\it{\fam\itfam\twelveit}
  \def\sl{\fam\slfam\twelvesl}     \def\bf{\fam\bffam\twelvebf}
  \def\mit{\fam 1}                 \def\cal{\fam 2}
  \def\tt{\twelvett}
  \textfont0=\twelverm   \scriptfont0=\tenrm   \scriptscriptfont0=\sevenrm
  \textfont1=\twelvei    \scriptfont1=\teni    \scriptscriptfont1=\seveni
  \textfont2=\twelvesy   \scriptfont2=\tensy   \scriptscriptfont2=\sevensy
  \textfont3=\twelveex   \scriptfont3=\twelveex  \scriptscriptfont3=\twelveex
  \textfont\itfam=\twelveit
  \textfont\slfam=\twelvesl
  \textfont\bffam=\twelvebf \scriptfont\bffam=\tenbf
  \scriptscriptfont\bffam=\sevenbf
  \normalbaselines\rm}
\def\IR{{\rm I\!R}}  
\def\IZ{\relax\ifmmode\mathchoice
{\hbox{\cmss Z\kern-.4em Z}}{\hbox{\cmss Z\kern-.4em Z}}
{\lower.9pt\hbox{\cmsss Z\kern-.4em Z}}
{\lower1.2pt\hbox{\cmsss Z\kern-.4em Z}}\else{\cmss Z\kern-.4em Z}\fi}

\def\oneandahalfspace{\baselineskip=\normalbaselineskip \multiply
\baselineskip by 1}

\newcount\equationnumber
\advance\equationnumber by1
\def\ifundefined#1{\expandafter\ifx\csname#1\endcsname\relax}
\def\docref#1{\ifundefined{#1} {\bf ?.?}\message{#1 not yet defined,}
\else \csname#1\endcsname \fi}
\def\autoeqnum{\def\eqlabel##1{\edef##1{\the\equationnumber}}}
\def\no{\eqno(\the\equationnumber){\global\advance\equationnumber by1}}
\newcount\citationnumber
\advance\citationnumber by1
\def\ifundefined#1{\expandafter\ifx\csname#1\endcsname\relax}
\def\cite#1{\ifundefined{#1} {\bf ?.?}\message{#1 not yet defined,}
\else \csname#1\endcsname \fi}
\def\autocite{\def\citelabel##1{\edef##1{\the\citationnumber}\global\advance\citationnumber by1}}
\def\preprintno#1{\rightline{\rm #1}}

\hsize=6.5truein
\hoffset=.1truein
\vsize=8.9truein
\voffset=.05truein
\parskip=\medskipamount
\twelvepoint           
\oneandahalfspace
\autocite
\autoeqnum 

\def\s{\scriptstyle}
\def\ss{\scriptscriptstyle}
\def\cl{\centerline}
\def\ds{\displaystyle}

\def\slashxi{{\xi}\!\!\!/} 
\def\nx{n(\xi,t)}

\def\o{\over}
\def\C{{\s C}}
\def\xL{\xi_{\ss L}}
\def\xR{\xi_{\ss R}}
\def\pmic{P(\C_i,t)}

\def\pmicl{P(\C_i,t+1)}

\def\pmicf{P'(\C_i,t)}
\def\pmicjf{P'(\C_j,t)}
\def\pmicc{P_c(\C_i,t)}
\def\pmicjc{P_c(\C_j,t)}

\def\pmicLf{P'(\C_i^{\ss L},t)}
\def\pmicRf{P'(\C_i^{\ss R},t)}
\def\px{P(\xi,t)}
\def\pxl{P(\xi,t+1)}

\def\pxf{P'(\xi,t)}
\def\pnxf{P'(\slashxi_i,t)}
\def\pxc{P_c(\xi,t)}
\def\pnxc{P_c(\slashxi_i,t)}

\def\pxLf{P'(\xi_{\ss L},t)}
\def\pxRf{P'(\xi_{\ss R},t)}
\def\pxL{P(\xi_{\ss L},t)}
\def\pxR{P(\xi_{\ss R},t)}

\def\fav{{\bar f}(t)}
\def\fxi{{\bar f}(\xi,t)}
\def\frat{{\fxi\over \fav}}
\def\fx{{f_{\xi}}}
\def\fxL{{f_{\xi_L}}}
\def\fxR{{f_{\xi_R}}}
\def\fmic{f(\C_i,t)}
\def\fmicj{f(\C_j,t)}

\def\dfx{\delta\fx}
\def\dfxL{\delta\fxL}
\def\dfxR{\delta\fxR}
\def\dfxi{\delta\fxi}

\def\str{{\cal A}_s}
\def\muti{{\cal P}({\ss C_i})}
\def\mutij{{\cal P}({\ss C_i\ra C_j})}
\def\mutji{{\cal P}({\ss C_j\ra C_i})}
\def\mutis{{\cal P}({\s \xi})}
\def\mutijs{{\cal P}({\s \slashxi_i\ra\xi})}
\def\crossij{{\cal C}_{\ss C_iC_j}^{(1)}(k)}
\def\crossjl{{\cal C}_{\ss C_jC_l}^{(2)}(k)}
\def\hamij{d^H(i,j)}
\def\hamijL{d^H_L(i,j)}
\def\hamijR{d^H_R(i,j)}
\def\hamilL{d^H_L(i,l)}
\def\hamilR{d^H_R(i,l)}
\def\ra{\rightarrow}

\def\ef{f_{\ss\rm eff}(\xi,t)}
\def\delef{\delta\ef}
\def\e{{\rm e}}

\vskip -48 truept
\preprintno{ICN-UNAM-96-08}
{ \hfill 30th October, 1996}
\break
\vskip 0.6truein


\cl{\bigg {ANALYSIS OF THE EFFECTIVE DEGREES OF}}
\vskip 0.3truein
\cl{\bigg {FREEDOM IN GENETIC ALGORITHMS}}
\vskip 0.3truein
\cl{\bgg C. R. Stephens}
\cl{and} 
\cl{\bgg  H. Waelbroeck}
\vskip\baselineskip
\cl{\it Instituto de Ciencias Nucleares, UNAM,}
\cl{\it Circuito Exterior, A.Postal 70-543,}
\cl{\it M\'exico D.F. 04510.}
\vskip 1truein

\noindent{\bf Abstract:}\ \ An evolution equation for a population of strings evolving
under the genetic operators: selection, mutation and crossover is derived.
The corresponding equation describing the evolution of schematas is found by
performing an exact coarse graining of this equation. In particular 
exact expressions for schemata reconstruction are derived which allows for
a critical appraisal of the ``building-block hypothesis'' of genetic algorithms. 
A further coarse-graining is made by considering the contribution of all length-$l$ 
schematas to the evolution of population observables such as 
fitness growth. As a test function for investigating the emergence of structure in the
evolution the increase per generation of the {\it in-schemata fitness} 
averaged over all schematas of length $l$, $\Delta_l$, is introduced.
In finding solutions of the evolution equations we concentrate 
more on the effects of crossover, in particular we consider crossover in the context
of Kauffman $Nk$ models with $k=0,2$. 
For $k=0$, with a random initial population,
in the first step of evolution the contribution from schemata reconstruction is equal to that
of schemata destruction leading to a scale invariant situation where the contribution
to fitness of schematas of size $l$ is independent of $l$. This balance is broken in
the next step of evolution leading to a situation where schematas that are either 
much larger or much smaller than half the string size dominate over those 
with $l \approx N/2$.
The balance between block destruction and reconstruction is also
broken in a $k>0$ landscape. It is conjectured that the effective 
degrees of freedom for such landscapes are 
{\it landscape connective trees} that break down into effectively fit 
smaller blocks, and not the blocks themselves. Numerical simulations 
confirm this ``connective tree hypothesis'' by showing that correlations 
drop off with connective distance and not with intrachromosomal distance.

\vfil \eject

\line{\bgg 1. Introduction \hfil}

One of the most important steps in developing a qualitative or 
quantitative model of a system is to gain an understanding
of the nature of the effective degrees of freedom of the system. This
is equally true if one is considering static, equilibrium properties or
dynamics; in the context of ``simple'' systems or of complex systems. 
An important feature that distinguishes the effective degrees of freedom
is that their mutual interactions are not very strong; that is to say that
they must have a certain degree of integrity. In this sense, the aim of 
developing an effective model of a system is to arrive at a 
description of the system in terms of {\it relevant} (e.g., ``macroscopic'')
variables.

 Identifying the correct effective degrees of freedom in complex
systems is generally speaking a very difficult task. To begin with, more often 
than not the effective degrees of freedom are scale dependent, where 
what one means by ``scale'' depends on the particular problem under 
consideration. In the case of evolution theory and genetic algorithms, 
one expects to find different effective degrees of freedom at different 
time scales. 
Generically if a system is complex at the relevant scale
then it will admit a simple effective dynamics only in 
terms of complex degrees of freedom: one trades the complicated dynamics
that results from the non-linear interactions 
of the many ``elementary" degrees of freedom for the simpler 
dynamics of more complicated effective degrees of freedom. 
What one gains in the trade is effective predictability; what one loses is detail. 

 It is well worth recalling in this context the example of spin glass models
of neural networks
\citelabel{\hopfield}\citelabel{\amit}\citelabel{\zertuche}
[\cite{hopfield}, \cite{amit}, \cite{zertuche}]. In this case
the effective degrees of freedom are the overlaps with a certain number 
of ``patterns'', each of which is related to a local extremum of the 
energy landscape or Hamiltonian. Since a large number of uncorrelated patterns 
is involved in this effective representation it should be clear that the
description of the effective degrees of freedom themselves requires a
large amount of information: One gets a 
measure of the complexity of the system by the information in its effective 
degrees of freedom. Note that in this example the system's dynamics is 
guided by large-scale attracting structures (the patterns), the effective
degrees of freedom (overlaps) being the instruments which measure how  
structure emerges as the system condenses from a disordered phase. 

 Not all complex systems have a large-scale structure which can be 
described in terms of macroscopic variables with a ``simple'' effective 
dynamics. For example in a critical sandpile \citelabel{\bak}[\cite{bak}], 
the relevant macroscopic variable is the avalanche size, and nothing short of a 
detailed description of every grain of sand would allow one to predict 
the size of the next event. Some examples of structured complex 
systems besides the Hopfield models include the brain, gene expression 
in eukaryotic cells \citelabel{\genesV}[\cite{genesV}], and of course 
evolution theory and genetic algorithms, among many others. 
We know that these systems are {\it structured} because their behaviour 
is manifestly non-random; for instance neural dynamics must be 
structured if the brain is to be of any use! Yet in most cases we have 
no idea what the nature of this structure {\it is}, much less how to 
identify effective degrees of freedom.

In this paper we will begin to analyse the notion of effective degree of 
freedom in the context of genetic algorithms (GA's) \citelabel{\holland}
\citelabel{\goldberg} [\cite{holland},\cite{goldberg}]. 
The claim that genetic algorithms are structured complex
systems is central to their designer's purpose, in that they yield 
intelligent solutions to complex optimization problems rather than 
a sophisticated sort of random search. We emphasize however 
that GA's form only one area of interest where the results and conclusions 
of this paper are applicable, some others being statistical mechanics 
\citelabel{\peliti} [\cite{peliti}] and biology [\cite{genesV}], the Kauffman $Nk$ 
model \citelabel{\kaufman}[\cite{kaufman}], and evolution 
theory \citelabel{\kimura}[\cite{kimura}]. 

 Trying to ascertain what effective
degrees of freedom a GA is using in order to arrive at an optimal solution is
in the strict sense a nonsensical question --- roughly equivalent to asking
``what are the effective degrees of freedom of a block of material?'' Of course,
the answer depends on the type of material under consideration and its state, 
the effective degrees of freedom of a superconductor being quite different to 
those of a spin-glass for instance. However, it is {\it not} non-sensical to think of
what are the effective degrees of freedom in a generic type of fitness landscape.
The fitness landscapes we choose to consider as being representative
of general classes of fitness landscapes are Kauffman's $Nk$ models with 
$k=0$ and $k=2$. 

 As in the example of spin glasses, the dynamics of genetic algorithms 
can be viewed as a condensation process in a rugged landscape. So 
again one expects the effective degrees of freedom to represent 
the emergence of certain structures, or ``patterns'', which are 
related to local fitness optima.
In GA theory one usually considers partly-specified patterns, called 
``schematas'', and determines the fraction of all the individuals in 
the population which include a particular schemata, this being a measure
of order comparable to the ``overlap'' of spin glass models. Since one does
not know a priori {\it which} schematas lead to a useful set of effective 
degrees of freedom some hypothesis must be made to this effect.

The standard conjecture about the effective degrees of freedom of genetic
algorithms is the ``building block hypothesis" [\cite{holland},\cite{goldberg}],
the essence of  which is that a GA arrives at an optimal solution of 
a complex problem via the combination of {\it short}, fit schematas. 
In this paper we will present both analytic and numerical evidence that 
generically this is not the case.
The argument for the block hypothesis is that large schematas 
are likely to be ``broken" by the crossover operator. Roughly speaking, 
the probability that a parent pass a length-$l$ schemata down to its offspring
drops off like ${l-1 \over N-1}$, where $N$ is the size of the string. 
However, this argument neglects the possibility that a schemata be repaired, if
the other parent has the part of the schemata that was broken by crossover;
more importantly as it turns out, it neglects the possibility that a
schemata be reconstructed from two parents that have incomplete parts of it.

It is clear that the validity of the block hypothesis will depend on the nature of
the fitness landscape. If there is a larger contribution to fitness from string bits
that are widely separated then clearly large schematas will be favoured irrespective
of the effect of crossover. On the contrary if the fitness landscape strongly favours
smaller schemata this would lend support to the block hypothesis. However, the 
intuition behind the block hypothesis is firmly based on the action of crossover not
with the pathologies of particular landscapes. It is for this reason that we
choose to consider the block hypothesis in the context of Kauffman $Nk$ models.
In particular, the case $k=0$ corresponds to the neutral case where
there are no bit-bit interactions to induce size dependence. In all cases, 
we will assume that the fitness landscape is generic in the sense that 
there is no systematic bias in the fitness function that would favour 
one part of the string over another. 

The bulk of this paper is devoted to deriving an equation for the
evolution of schematas, and from there a coarse-grained equation
which gives the average contribution of schematas of size $l$ to the 
improvement of fitness. We will show that under general assumptions
the coarse-grained variable is closely related  to the spatial correlation 
function, so it provides information about the size distribution of 
the effective degrees of freedom.
We will apply this equation to particular situations, to analyse
the effect of crossover on the emergence of correlations between distant 
bits in the strings. Both the theoretical analysis and the numerical 
simulations lead to a new conjecture about the  effective degrees 
of freedom of direct-encoded genetic algorithms on an $Nk$-landscape,
which we call the ``connective tree hypothesis''. 

The format of the paper will be as follows: in section 2, as this paper 
is not intended for a dedicated GA audience, we will give a brief overview 
of various elements of GA theory. In section 3 we will derive an 
evolution equation for the development of a population of strings
under the genetic operators of selection, mutation and simple crossover. 
We then ``coarse grain'' this equation to derive an effective evolution 
equation for the evolution of schematas of size $l$ and order $N_2$, 
thus arriving at a generalization of the fundamental theorem of 
schematas [\cite{holland}]. In section 4 we consider a further coarse 
graining considering the effects of schematas of size $l$ but of any 
order $N_2\leq l$. We consider especially the increment in fitness per 
generation from such schematas. In section 5 we consider asymptotic 
solutions of the coarse grained evolution equation near a random initial 
population for a simple ``neutral'' fitness landscape
and also make some comments about what happens near the ordered 
population limit. In section 6 we consider a more non-trivial landscape 
--- a Kauffman $Nk$ model [\cite{kaufman}] with $k=2$. Finally in section 
7 we summarize our conclusions.

\vskip 0.3truein 

\line{\bgg 2. Genetic Algorithms and the Building Block Hypothesis \hfil}

GA's have become increasingly popular in the analysis of complex search and
optimization problems and in machine learning, one of their chief attributes 
being their robustness (see \citelabel{\mitchell}[\cite{mitchell}] and references
therein for a recent overview). 
One begins with a complex optimization problem which depends on many 
variables. The variables and the rules that govern them are subsequently coded
in the form of a population of strings/``chromosomes''. The latter consist of a set
of  symbols/``alleles'', each symbol taking values defined over an alphabet. 
Here we will only consider
binary codification though our general conclusions apply also to alphabets
of higher cardinality. We will denote by ${\cal A}_s$ the space of possible 
states of a string. 

The population is evolved under
the action of a set of genetic operators. Reproduction can be implemented 
in many different ways, all have the
effect of increasing the relative numbers of ``fit'' strings between 
one generation 
and another; fitness being measured by a fitness function, $f:\str\ra \IR_+$. 
The role of most other genetic operators is to encourage diversity in
the population. In this paper we will restrict our attention to simple crossover and  
mutation. The former is a type of recombination and involves the splitting of 
two parents, $C_i, C_j \in {\cal A}_s$, at a particular crossover point $k$, 
and the subsequent juxtaposition and recombination of the left half of 
$C_i$ with the right half of $C_j$ and the right half of $C_i$ with the left 
half of $C_j$, left and right being defined relative to the crossover point
$k$. As mentioned, the point of genetic operators such as crossover is 
to encourage population diversity. Optimal strings that do not appear in
the initial population cannot subsequently appear through the effects of reproduction alone.
Crossover is one method for generating fit strings that weren't originally
in the population of a given generation. Mutation on the other hand offers 
a form of insureance in that if a particular bit is lost it is 
irrecoverable using only reproduction and crossover. Mutation
offers a way to recover lost bits that may subsequently be important in the construction
of an optimum string. Speaking intuitively, we may say that relative to an optimum string, 
mutation produces ``errors''  whereas crossover merely shuffles them around. We will find
this distinction to be an important one when we come to critique the building block hypothesis
later. Using the language of statistical mechanics the evolution of the GA is 
a competition between the ``ordering'' tendency of reproduction and the ``disordering'' effects
of crossover and mutation. 

The language used in GA theory obviously owes much to 
evolutionary biology, indeed the whole point of GA's was to try to adapt the methods used by 
adaptive systems in nature in the context of artificial systems; selection, mutation and
 recombination being extremely important elements in the search for ``fit'' organisms. In the
discussion above, for instance, a string could represent a protein chain of a certain size and 
the possible values of a symbol the number of possible alleles at a particular site
on the chain.

Theoretical analysis of how a GA seeks an optimum solution has focussed
on the notion of schematas. If we consider strings of $N$ bits, a schemata is a subset, 
$N_2\leq N$, of bits defining a certain ``word'' constructed from the 
alphabet. In the $N-N_2$ positions not defined by the schemata one does 
not care about the value of the bit and this is taken into account 
by use of the metasymbol, or ``wildcard'', $\star$. For an alphabet of size $m$ there 
are $\alpha_{\ss\xi}=(m+1)^N$ possible schematas for a particular string.
The total number of possible schematas in the population is 
$n_{\ss\xi}\leq n(m+1)^N$, the exact number depending on the 
population diversity. For a totally organised population 
$n_{\ss\xi}=\alpha_{\ss\xi}$. 

The essential idea behind the notion of schemata is that the GA arrives 
at an optimum solution through combining fit schematas. As each string is
an example of $\sim 2^N$ schematas it is clear that a very large number of them are
being processed simultaneously by the GA, a phenomenon known as implicit
parallelism [\cite{holland}]. Of course, not all these schematas survive crossover, 
which leads us to consider
the size of a schemata, $l$, which is defined as $(l = j-i+1)$, where $i$ and $j$ are the first and last
of the $N_2$ defining elements of the schemata respectively. In terms of reproduction alone
there is no preference for short versus long schematas, except as might be induced by the 
fitness function itself. Equally, mutation shows no favour
for one or the other. However, if one considers the effects of crossover, purely in terms of the
crossover point itself there is a higher probability to ``break'' a long schemata than a short one.
This of course neglects the possibility of reconstructing a schemata even though it does not
appear in either of the parents involved in the crossover process. This apparent disfavour for
large schematas imposed by crossover has led to what is known as the ``building block'' 
hypothesis which claims that the joint effect of reproduction and crossover is to favour highly
fit but {\it short} schematas which propagate from generation to generation exponentially. It is these highly fit, short schematas which are then
considered to be the effective degrees of freedom in the system, the GA
building a better solution through combining small sub-solutions. 

\vskip 0.3truein

\line{\bgg 3. String Evolution Equation \hfil}

In this section we will derive an equation that describes the evolution of a 
GA induced by the effects of the three genetic operators: selection,
crossover and mutation. In particular we will consider the change in number 
$\nx$ of strings that contain a
particular schemata $\xi$, of order $N_2$ and size $l\geq N_2$, as a function of 
time (generation) in a population of size $n$. 
It is worth pointing out here that a schemata itself is already a coarse grained 
degree of freedom in the sense that to calculate any properties of a schemata, 
such as its fitness, one needs to take a population average. 

We will first derive evolution equations for the ``microscopic'' degrees of freedom themselves
--- the strings. Considering first selection in the absence of mutation or crossover one has
\eqlabel{\mast}
$$\pmicl=\pmicf\no$$
where $\pmicf=(\fmic/\fav) P(C_i,t)$,
$\fmic$ is the fitness of string $\C_i$ at time $t$, $\pmic=n(\C_i,t)/n$ and
$\fav=\sum_i\fmic\pmic$ is the average string fitness. 
In (\docref{mast}) we are neglecting fluctuations in the numbers
$ n(C_i,t)$, an approximation which should be reasonable as long as the population
is not too ``sparse", we will return to this point later.
Clearly, as previously mentioned, the effect of reproduction is to augment the number
of fit strings, fit here meaning $\fmic>\fav$, and to decrease the number of unfit strings, where
by unfit we mean $\fmic<\fav$. In terms of reproduction alone
it is simple to prove that average fitness is a Lyapunov function, increasing monotonically 
as a function of time. As also discussed in the last section, the trouble with 
using selection as the sole genetic operator is that the search space for optima is
restricted to that of the initial population. In complex systems this number will usually be
negligible compared to the size of the total search space. This implies that
finite size effects are important. The theory of branching processes may be a more 
suitable framework in this regard \citelabel{\taib}[\cite{taib}] . 
 
Including in the effects of mutation but not crossover gives rise to the quasi-species model
\citelabel{\eigen}[\cite{eigen}], with evolution equation
\eqlabel{\mastmut}
$$\pmicl=\muti\pmicf+{\ds\sum_{\ss C_j\neq C_i}}\mutji\pmicjf\no$$
where $\muti=\prod_{\s k=1}^{\ss N}(1-p(k))$ is the probability that 
string $i$ remains unmutated, $p(k)$ being the probability of mutation 
of bit $k$ which we assume to be a constant, though the equations are 
essentially unchanged if we also include a dependence on time. 
$\mutji$ is the probability that string $j$ is mutated into string $i$,
$$\mutji=\prod_{\ss{k\in \{C_j-C_i\}}}{p(k)}\prod_{\ss{k\in\{C_j-C_i\}_c}}
{(1-p(k))}\no$$
where $\s\{C_j-C_i\}$ is the set of bits that differ between $\C_j$ and 
$\C_i$ and $\s\{C_j-C_i\}_c$, the complement of this set, is the set 
of bits that are the same. In the limit where the mutation
rate $p$ is uniform, $\muti=(1-p)^{\s N}$ and 
$\mutji=p^{\s d^{\ss H}(i,j)}(1-p)^{\s N-d^{\ss H}(i,j)}$,
where $\hamij$ is the Hamming distance between the strings $\C_i$ and $\C_j$.
The behaviour of the solutions of equation (\docref{mastmut}) has been much 
discussed in the literature (see for example [\cite{peliti}] 
and references therein), although mainly in the context of a flat fitness 
landscape. One of the principal features of interest is the existence 
of an ``error threshold'' separating an ``ordered'' (selection dominated) 
phase from a ``disordered'' (mutation dominated) phase which manifests 
itself as a second order phase transition at a certain critical mutation rate.

We will now consider the effects of crossover without mutation. This is a
much less studied case theoretically (though see \citelabel{\benn}[\cite{benn}]), 
but one that is very important from
the point of view of effective degrees of freedom, since unlike mutations
it is sensitive to the linear disposition of bits along the string. It is also plays a
very important role in biology.
With crossover the evolution equation can be written in the form
\eqlabel{\mastcross}
$$\eqalign{\pmicl&=\pmicf-{p_c\o N-1}{\ds\sum_{\ss C_j\neq C_i}}{\ds\sum_{k=1}^{\ss N-1}}
\crossij\pmicf\pmicjf\cr 
&+{p_c\o N-1}\sum_{\ss C_j\neq C_i}\sum_{\ss C_l\neq C_i}\sum_{k=1}^{\ss N-1}
\crossjl\pmicjf P'(\C_l, t)
\cr}\no$$
where $p_c$ is the probability to implement crossover in the first place,
\eqlabel{\thetas}
$$\crossij=\theta(\hamijL)\theta(\hamijR)\no$$
and
\eqlabel{\deltas}
$$\crossjl={1\o2}\left(\delta(\hamijL)\delta(\hamilR)+\delta(\hamijR)\delta(\hamilL)\right)\no$$
where $\hamijR$ is the Hamming distance between the right halves of the strings $C_i$ and
$C_j$,  ``right'' being defined relative to the crossover point $k$. The other quantities are defined analogously. 
$\crossij$ is the probability that given that $\C_i$ 
was one of the parents it is destroyed by the crossover process. 
$\crossjl$ is the probability that given that neither parent was $\C_i$ it 
is created by the crossover process, so this represents
a gain term. It is naturally much easier 
to destroy an individual string by crossover than create it hence $\crossjl$ 
is a very sparse matrix.  $\crossjl$ represents a contact interaction term in 
Hamming space. Another important property of $\crossij$ and $\crossjl$ is that they are 
completely population independent, depending only on string configurations and not string
numbers.

In the case of mutations and selection without crossover 
the non-equilibrium evolution equation has been mapped into an equilibrium 
statistical mechanics problem using transfer matrix techniques 
\citelabel{\statan} [\cite{statan}], where the role of
inverse temperature is played by $\beta={1\o2}{\rm log}(p/(1-p))$. 
One can also find an analogy for the crossover operator which can 
provide a more intuitive understanding of its effects. Imagine a ``population'' of $n$ one-dimensional Ising chains in a strong magnetic field 
$h$, where the effects of spin-spin couplings may be neglected. We denote spin up by $1$ and spin down by $0$. 
The ``fitness'' of string $i$ is simply $f(\C_i)=h (n_1-n_0)$. Clearly
selection will favour strings with large values of $n_1$ relative to $n_0$. So what are the effects
of crossover? Consider two selected chains of size $N=7$: $1111000$ and $0100111$. One
may think of the first chain as being composed of a domain of up spins of size 
$4$ and a domain of
down spins of size $3$ separated by a domain wall or ``kink". Similarly, the second chain 
consists of two domains of up spins of sizes $1$ and $3$, and two of down spins of sizes $1$ and $2$. 
These domains are separated by two kinks and two anti-kinks. If the crossover occurs 
at $k=5$ the resultant chains are $1111111$ and $0100000$. Chain number one is now 
``homogeneous'', containing no kinks or anti-kinks, whilst chain two contains a 
kink--anti-kink pair. Thus the action of crossover has been the annihilation of a kink--anti-kink 
pair. A crossover at $k=3$ would have yielded: $1100111$ and $0111000$, each chain now 
having one kink--anti-kink pair. One can think of this as just a kink--kink scattering process
 that has conserved the total number of kinks and anti-kinks. Finally,  a crossover at $k=6$ would
give $1111011$ and $0100100$ where now there are three kink--anti-kink pairs, one in the first chain 
and two in the second. In this case there has been kink--anti-kink creation. 
Thus we see here that crossover may be interpreted as creation, annihilation 
and scattering of kinks --- domain walls, thought of as topological defects. Here, the fitness 
landscape is such as to favour
spin chains without topological defects. This is because we are considering a ``ferromagnetic''
fitness landscape. If the fitness landscape were such as to favour 0's in odd positions and 
1's in even positions then an optimum chain would be $0101010$, i.e. an inhomogeneous state with a maximum number of kink--anti-kink pairs. Generically then, crossover may be
thought of as inducing interactions between ``domains''. Exactly 
what type of domain is favoured is 
of course a function of what is the particular fitness function of interest. 
It is also worth noting that in the population crossover without selection preserves the 
total number of  0's and 1's in any given bit position in the population. Thus for instance, if
we consider a non-optimum string as having ``errors'' relative to an optimum string, pure
crossover without selection cannot change the total number of errors in the population 
it can only shuffle them around.

Equation (\docref{mastcross}) is an extension of the 
``schema theorem'', or fundamental theorem of GA's, [\cite{holland},\cite{goldberg}] 
which states that for a schemata, $\xi$, of size $l$
\eqlabel{\schth}
$$\pxl\geq\pxf\left(1-p_c\left({l-1\o N-1}\right)\right),\no$$
to the case where the schemata of interest is the entire string (an 
analogous equation was derived in \citelabel{\goldbridge} [\cite{goldbridge}]). 
The evolution equation we have derived takes into account 
exactly, given the approximation of a large population, the effects of destruction 
and reconstruction of strings. 

Combining the effects of both crossover and mutation, where we assume that mutation is 
carried out after crossover, we have the evolution equation
\eqlabel{\mastcrossmut}
$$\pmicl=\muti\pmicc + {\ds\sum_{\ss C_j\neq C_i}}\mutji\pmicjc$$
where
$$\eqalign{\pmicc &= \pmicf-{p_c\o N-1}{\ds\sum_{\ss C_j\neq C_i}}{\ds\sum_{k=1}^{\ss N-1}}
\crossij\pmicf\pmicjf\cr 
&+{p_c\o N-1}\sum_{\ss C_j\neq C_i}\sum_{\ss C_l\neq C_i}\sum_{k=1}^{\ss N-1}
\crossjl\pmicjf P'(\C_l, t)
\cr}\no$$

The various evolution equations exhibit different dynamical fixed points. 
In the ``low temperature limit'' $p \to 0$, in a non-trivial fitness landscape, one has 
stable fixed points at the local fitness extrema. If $\fmic>\fmicj\ \forall \C_j
: \mutji > 0$, then
$$n(\C_i)=n, \ \ n(\C_j)=0 \no$$
is a stable fixed point of (\docref{mastcrossmut}).
These evolution equations are exact only in the approximation of an infinite 
population. For a finite population there are no stable fixed points (for $p\neq 0$)
due to the effect of fluctuations in the reproduction process.
This can be seen in the simple example of two types of string, A and B, with a 
population of size $10$ and fitnesses $f_A$ and $f_B$ $(f_A>f_B)$. If 
we start with equal proportions of A and B the probability that 
$n_A\ra6$ relative to the probability that $n_A\ra4$ is $(f_A/f_B)^2$. 
So the effect of fitness is to increase the probability that we have 
more A's in the population. Consider now if the initial population is 
$n_A=1$, $n_B=9$. The probability that A will disappear at the first 
time step is $S_A=(1+(f_A/9f_B))^{\s -10}$.  In a flat fitness landscape
$S_A=0.35$. For $f_A=2f_B$, $S_A=0.13$ and for $f_A=10f_B$, $S_A=0.01$. 
We see then that unless the fitness advantage of A over B is quite 
pronounced there is a non-trivial probability that the fitter fixed 
point is not reached, whereas reaching the latter is an inevitable conclusion 
of the evolution equation neglecting fluctuations. In general, the Neutral 
Theory predicts that the selective coefficient must be greater than $1/n$ 
to ensure that selection dominates over random drift.

We now turn our attention to the derivation of an evolution equation for 
schematas, $\xi$, of order $N_2$ and size $l$. Before doing this it is convenient
to return to equation (\docref{mastcross}) to see that the notions of schemata
and coarse graining appear very naturally when considering crossover of
strings. Considering the destruction term: the matrix (\docref{thetas}) 
restricts the sum to those $C_j$ that differ from $C_i$ in at least one bit both
to the left and to the right of the crossover point. One can convert the sum over
$C_j$ into an unrestricted sum by subtracting off those $C_j$ that have 
$d^H_L(i,j)=0$ and/or $d^H_R(i,j)=0$. Similarly one may write the 
reconstruction term as
\eqlabel{\reconterm}
$$\eqalign{{p_c\o N-1}\sum_{k=1}^{N-1}&\left(\sum_{\ss C_j}\sum_{\ss C_l}\crossjl
\pmicjf P'(\C_l, t)\right.\cr
&\left.-2\sum_{\ss C_j}{{\cal C}_{\ss C_iC_j}^{(2)}(k)}
\pmicf \pmicjf-\pmicf\pmicf\right)\cr}\no$$
The second and third terms cancel with corresponding expressions from the 
destruction term hence (\docref{reconterm}) can now be written as
$$\sum_{\ss C_j\supset C_i^L}\sum_{\ss C_l\supset C_i^R}
\pmicjf P'(\C_l, t)\no$$
where $C_i^L$ is the part of $C_i$ to the left of the crossover point and
corespondingly for $C_i^R$. However, by definition
$${\bar f}(C_i^{\ss L},t)={1\o n_{\ss C_i^L}}\sum_{\ss C_j\supset C_i^L}\pmicjf\no$$
where $n_{\ss C_i^L}$ is the total number of strings in the population 
that contain $C_i^L$. As ${\bar f}(C_i^L,t)$ is the average fitness of the 
substring $C_i^L$, one can think of this substring as a schemata,
likewise for $C_i^R$. In terms of these ``schematas'' the final form 
of the string equation is
\eqlabel{\stringfin}
$$\pmicl=\pmicf -{p_c\o N-1}\sum_{k=1}^{\s N-1}(\pmicf-\pmicLf\pmicRf)\no$$
with 
$$\pmicLf=\sum_{\ss C_j\supset C_i^L}\pmicjf\no$$
and similarly for $\pmicRf$. 

One sees that crossover very naturally introduces
the notion of coarse graining even though we are working in terms of the microscopic
degrees of freedom --- the strings. The reconstruction probability depends on the
relative fitness of strings that contain the constituent elements of $C_i$, but given that
there can be many strings that contain $C_i^L$ one must take an 
average over these strings. In this sense we are integrating out the ``degrees of 
freedom'' represented by the bits that are not contained in $C_i^L$ or $C_i^R$.
Equation (\docref{stringfin}) shows that the 
effects of reconstruction will outweigh destruction if the parts of a string are 
more selected than the whole.

Before deriving a schemata evolution equation including crossover and
mutation let us consider the effects of reproduction alone. The 
proportion of elements of the 
population, $\px$, that contain $\xi$ satisfies the evolution equation
\eqlabel{\evorep}
$$\pxl=\pxf\no$$
where $\pxf=\frat\px$, 
$$\fxi={\ds{\sum_{\ss{C_i\supset \xi}}^{\nx}}f( \C_i,t) n(\C_i,t)\o \nx}, \no$$ 
the sum is over all strings $\C_i$
that contain $\xi$, and $\fav=\sum_{\xi}\fxi\px/ \sum_{\xi}\px$ is the 
average fitness per string or per schemata of the population. 
Note that the sum over 
strings that contain $\xi$ is a sum over the possible values of the
bits that are not definite elements of $\xi$, i.e. the wildcards. In this
sense, as above in (\docref{stringfin}), ``degrees of freedom'' have 
been integrated out of the problem
and (\docref{evorep}) represents an exact coarse-graining of the 
original string evolution equation.

Note that the sum over different binary words on the $N_2$
defining bits of the schemata is a partition of the identity, i.e.
$$\sum_{\ss{\rm words}}\px=1\no$$
If we sum also over different possible schemata configurations we have
$$\sum_{\xi}1=\sum_{N_2=1}^N{\ }^NC_{N_2}\sum_{\ss{\rm words}}1=3^N-1\no$$
which is just the total number of schematas (except for the order 0 schemata with no
defining bits). 

Considering mutation without crossover we ``coarse grain'' the microscopic equation
(\docref{mastmut}) by summing over all $\C_i\supset\xi$. One can write an effective
evolution equation for schematas evolving under mutation
\eqlabel{\mastmutsch}
$$\pxl=\mutis\pxf + \sum_{\s \slashxi_i}\mutijs\pnxf\no$$
where the effective coefficients $\mutis$ and $\mutijs$ are 
$$\mutis=\prod_{k=1}^{N_2}(1-p(k)) \no$$
and
$$\mutijs={\ds{\sum_{\ss{C_j\supset \slashxi_i}}}\pmicjf\mutij\o
n(\slashxi,t){\bar f}(\slashxi,t)}\no$$
In the latter the sum is over strings $\C_j$ that contain the schemata 
$\slashxi_i$, where $\slashxi_i$ differs in at least one bit from $\xi$ on the
$N_2$ defining bits of the schemata.

As with strings the effect of recombination in the form of crossover 
is two fold: it potentially destroys
schematas that were present in one parent but not the other; on the other
hand it offers the possibility of reconstructing schematas even though they were not
present in either parent. 
To derive an evolution equation for schematas, including in the effects of crossover,
we return to equation (\docref{stringfin}) and sum over all strings $C_i\supset \xi$.
One finds
$$\pxl=(1-p_c)\pxf +{p_c\o N-1}\sum_{\ss C_i\supset\xi}\sum_{k=1}^{N-1}
\pmicLf\pmicRf\no$$
We now break the sum over crossover points into those that cut 
the schemata itself and those
that cut outside the schemata. In the reconstruction term if the cut is outside the 
schemata, to the right say, then the sum over $C_i^R$ is one. Similarly if the cut is 
to the left with the sum over $C_i^L$. The remaining sums yield $\pxf$ and this term 
cancels with an analogous expression originating in the destruction term. For 
the reconstruction contribution from cuts in the schemata we denote by $\eta_L$
($\eta_R$) the bits to the left (right) of the crossover point that are {\it not}
in the schemata and note that
$$\sum_{\ss C_i\supset\xi}\pmicLf\pmicRf=
\sum_{\ss \eta_L}\sum_{\ss \eta_R}\pmicLf\pmicRf.\no$$
We will denote by $\xi_L$ and $\xi_R$ the parts of the schemata to the left and right
of the crossover point respectively. Now, 
$\sum_{\ss \eta_L}\pmicLf=\pxLf$, where by definition
$\pxLf$ $=({\bar f}(\xi_L,t)/\fav)P(\xi_L,t)$, ${\bar f}(\xi_L,t)$ being the average fitness
of the schemata $\xi_L$. Analogous expressions hold for $\xi_R$. With these 
results the final form of the schemata evolution equation including crossover is
\eqlabel{\maseqone}
$$\pxl=\pxf-{p_c\o N-1}\sum_{k=1}^{l-1}\left(\pxf-\pxLf\pxRf\right)\no$$
where the sum is only over crossover points that cut the schemata. 

The interpretation of this equation is very similar to that of (\docref{stringfin}).
In the reconstruction term  $\pxLf\pxRf$ is the probability that one parent is
selected that contains the left part of the schemata and the other contains 
the right part. A schemata will be augmented by the effects of 
crossover if, as in the string case, its constituent parts are selected more than the
whole schemata.
Compared with (\docref{stringfin}) a further coarse graining has been carried 
out by summing over all the states of bits outside of $\xi$. 
Combining now the effects of selection, mutation and crossover the schemata
evolution equation is
\eqlabel{\maseqtwo}
$$\pxl=\mutis\pxc + \sum_{\s \slashxi_i}\mutijs\pnxc\no$$
where
$$\pxc= \pxf-{p_c\o N-1}\sum_{k=1}^{l-1}\left(\pxf-\pxLf\pxRf\right)\no$$
This evolution equation is the fundamental equation governing the evolution 
of schematas and is written at a ``semi-microscopic'' level in that 
it is written in terms of individual schematas. It represents an exact coarse-graining
of the corresponding string evolution equation after summing over all possible
states of the non-schemata degrees of freedom. 

Another useful concept we will introduce here is that of  ``effective fitness'', 
$\ef$, which we define via the relation
\eqlabel{\efffit}
$$\pxl={\ef\o\fav}\px\no$$
comparing with equation (\docref{maseqtwo}) one finds
\eqlabel{\efffitex}
$$\eqalign{\ef=&\mutis\left(1-p_{\ss c}\left({l-1\o N-1}\right)\right)\fxi
+\sum_{\s \slashxi_i}\mutijs{\pnxf\o\px}{\bar f}(\xi_i,t)\cr
&+p_{\ss c}\left({ l-1\o N-1}\right){\mutis\o\px}\fav
\pxLf \pxRf\cr
&-{p_{\ss c} \o N-1}\sum_{\s \slashxi_i}\mutijs
\fav \sum_{k=1}^{N-1} \left({P'(\slashxi_i, t)
-P'(\slashxi_{i_L}, t)P'(\slashxi_{i_R}, t) \o \px} \right)    \cr}\no$$
Thus we see that the effect of mutation and crossover is to ``renormalize''
the ``bare'' fitness $\fxi$. The destructive effects of crossover and mutation 
give a multiplicative type renormalization whilst the reconstruction terms 
give an additive type renormalization. In the low ``temperature'' limit 
where mutation and crossover go to zero $\ef\ra\fxi$. 

Another concept we will find useful is that of an effective selection 
coefficient $s_{\ss\rm eff}=\ef/\fav-1$. If we think of $s_{\ss\rm eff}$ 
as being approximately constant in the vicinity of time $t_{\ss 0}$, 
then $s_{\ss\rm eff}(t_0)$ gives us the exponential rate of increase or 
decrease of growth of the schemata $\xi$ at time $t_0$.

\vskip 0.3truein

\line{\bgg 4. Effective Degrees of Freedom and Coarse Graining \hfil}

As mentioned in the introduction one of the most important steps in
obtaining a qualitative and quantitative understanding of a system is
deciding what are the relevant degrees of freedom of the system. This is 
often a quite difficult thing to do owing to the fact that they 
are ``scale''  dependent. In the case of 
evolution dynamics this means that the effective dynamics depends on
the time scale considered. Trying to understand such time dependent
behaviour quantitatively is very difficult as almost inevitably one will
have to resort to an approximation technique, which invariably
depends on focusing on the relevant effective degrees of freedom as in 
the methods of effective field theory. However, if they 
are time dependent then what starts as a good approximation 
focusing on a certain  type at one time will usually
break down as one approaches time scales where they are 
qualitatively quite different.

One feature that is very common, if one has found a reasonable set 
of effective degrees of freedom, is that their mutual interactions 
are not very strong, so that they have a 
certain degree of integrity. Calling something 
an effective degree of freedom is not a very useful thing to do if it is
not readily identifiable as such. For instance, in low energy QCD  
gluons and quarks are not very useful concepts as they
are so strongly coupled via highly non-linear interactions that they 
form baryons and mesons, bound states of the former. The
latter have a much higher degree of integrity than the former at such energies.

So how is the above related to the present discussion of GA's?
GA's, as algorithmic representations of complex systems, have
many degrees of freedom. For instance, in the case where the state of a string
of size $N$ is defined as a binary word, for a population $n$ the total number
of possible states is $\sim (2^N)^n$ in the case where strings are 
identifiable by a label other than the state of their bits, and $\sim n2^N$ in the 
case where permutations of identical strings are not counted separately. Both
are exponentially large numbers. 

Ideally, the search for optima proceeds in a smaller space,
spanned by effective ``coarse-grained'' degrees of freedom. 
The traditional answer to the question: ``What is the nature of these 
degrees of freedom?''  is, as mentioned
previously, the ``building block hypothesis'': that
small segments of string have an activity that is relatively
decoupled from the rest of a string --- these ``blocks'' are
assumed to be sufficiently compact that they have a high probability
of being preserved by crossover. The GA supposedly uses 
these building blocks in order to arrive at a global solution.

We can get some idea of the dynamical behaviour of schemata due
to crossover by restricting attention for the moment to a flat fitness landscape. 
In this case 
\eqlabel{\evolflat}
$$\pxl=\px-{p_c\o N-1}\sum_{k=1}^{l-1}\left(\px-\pxL\pxR\right)\no$$
For an uncorrelated population crossover is completely neutral and 
we have a scale invariant situation.

To solve the evolution equation (\docref{evolflat}) in the case of a 
correlated population one needs to solve the corresponding equations 
for $\xL$ and $\xR$; these will involve reconstruction
terms that contain $\xi_{\ss LL}$, $\xi_{\ss LR}$, $\xi_{\ss RL}$ and $\xi_{\ss RR}$.
The first two are the components of $\xL$ and the latter two of $\xR$. Naturally this
process can be iterated relating fine grained degrees of freedom to more and 
more coarse grained degrees of freedom, where more and more bits $(N-N_2)$
have been summed over. Obviously when one arrives at one schematas, the maximally
coarse grained degrees of freedom, the process stops as one cannot split
by crossover such schematas. We see then that crossover leads to an 
hierarchy of equations relating fine grained degrees of freedom to successively more
and more coarse grained degrees of freedom. 

Restricting attention to two schematas in the flat fitness landscape setting and 
considering the continuous time limit one arrives at the following differential 
equation
$${dP(ij,t)\o dt}=-p_c{l-1\o N-1}\left(P(ij,t)-P(i,t)P(j,t)\right)\no$$
where $i$ and $j$ are the definite bits that define the two schemata and also
the two one schematas respectively. As one cannot split a one schemata $P(i,t)$
and $P(j,t)$ are conserved quantities thus one finds
$$P(ij,t)=P(ij,0)\e^{-p_c{l-1\o N-1}t}+P(i,0)P(j,0)\left(1-\e^{-p_c{l-1\o N-1}t}\right)\no$$
Thus one sees that $P(ij,t)$ approaches an uncorrelated fixed point 
$P^*(ij)=P(i,0)P(j,0)$ exponentially rapidly. The sole effect of the 
size of the schemata is to govern the rate of approach to the fixed 
point, an exponentially small preference being given to smaller schematas. 

The steady state solution for a schemata $\xi$ of order $N_2$ is
$$P^*(\xi)=\prod_{i=1}^{N_2}P(\xi(i),0)\no$$
where $P(\xi(i),0)$ is the probability of finding the one schemata corresponding to
the i'th bit of $\xi$ at $t=0$. One can verify that this steady state solution also is a result
purely of the effects of reconstruction. Without reconstruction there is no other fixed point 
other than zero! We see then that reconstruction is the driving force of crossover and will
always come to dominate. This is very much contrary to the standard block hypothesis 
point of view which treats schemata destruction as the dominant effect. We can also make
another interesting observation associated with the effective fitness $\ef$ and crossover.
Here the effect of crossover is to renormalize the fitness. The effective selection coefficient is
\eqlabel{\selecflat}
$$s_{\ss\rm eff}=-p_c\left({l-1\o N-1}\right)+p_c\left({l-1\o N-1}\right){P(i,0)P(j,0)\o P(ij,t)}\no$$
Thus schemata destruction gives a multiplicative renormalization that 
contributes negatively to the effective fitness advantage. However, 
schemata reconstruction leads to an additive renormalization of the 
effective fitness which exceeds the contribution of the destruction term
if $i$ and $j$ are negatively correlated.

In general the fitness landscape itself induces correlations between 
$\xi_L$ and $\xi_R$. In this case there is a competition 
between the (anti-) correlating effect of the landscape and 
the mixing effect of crossover. Selection itself more often than not 
induces an {\it anti}-correlation between fit schemata parts, rather than
a positive correlation. Indeed, in the neutral case of a $k=0$ landscape 
one has $1 + {2N_2 \o N} \dfx < (1 + {2N_L \o N} \dfxL)
(1 + {2N_R \o N} \dfxR)$, so selection induces an anticorrelation when 
$\dfxL, \dfxR > 0$: In an uncorrelated initial population, 
$P'(\xi, t) < P'(\xi_L, t) P'(\xi_R,t)$. This means that crossover 
plays an important role in allowing both parts of a successful schemata to
appear in the same individual. 

We can analyze this effect in more detail taking once again the case of 
2-schematas. Defining the correlation 
${\cal C}(ij, t) \equiv (P(ij, t)/P(i,t)P(j,t))-1$
then in terms of the selection coefficient, 
$s_{\xi} = {\bar f}(\xi, t) / {\bar f} - 1$, one finds 
$${\cal C}(ij, t+1) = \left(1-p_c{l-1\o N-1}\right)\left({(1+s_{ij})\o
(1+s_i)(1+s_j)}
{\cal C}(ij, t) - {(s_is_j+s_i+s_j)\o (1+s_i)(1+s_j)}\right)\no.$$
Note that the effect of crossover is to diminish correlations induced by
the fitness landscape, however crossover cannot change the sign of the
correlations. The larger the value of $l$ in this simple case
the more the correlations are damped.

This is the effect which we saw previously in the 
context of a flat landscape. In the extreme case $l=N$, $p_c=1$ the 
effect of crossover is to eliminate all correlation between $i$ and $j$. 
In the neutral ($k=0$) case, $s_{ij} = s_i + s_j$ and 
$${\cal C}(ij, t+1) = \left(1-p_c{(l-1)\o N-1}\right)
\left({(1+s_i+s_j)\o (1+s_i)(1+s_j)}{\cal C}(ij, t) - {s_is_j\o
(1+s_i)(1+s_j)}\right).\no$$
Thus the effect of crossover is to weaken but not cancel completely the
anti-correlations induced by $k=0$ selection. In the remainder of this section
we will consider this effect for general schematas.

In our search for the relevant effective degrees of freedom and in analysing 
the building block hypothesis we will consider schematas
of length $l$ irrespective of their order or their overall position in a string. It should
be clear that this is a further coarse graining relative to the evolution
equations considered earlier. Unfortunately the evolution equation 
(\docref{maseqtwo}) by itself is not very useful for 
analysing schematas of size $l$, the reason being that any given string contains
schematas of all sizes. However, consideration of just about any quantity in conjunction
with (\docref{maseqtwo}) and a sum over schematas of a given length is meaningful. 
For instance, one could consider how $\fxi$ changes in time and subsequently how its
average, $<\fxi>_l$, over all possible schematas of size $l$ changes. 
Our notation here is that for any function $A(\xi, t)$, 
\eqlabel{\Aav}
$$<A(t)>_l={{\ds{\sum_{i=1}^{\ss (N-l+1)}}}{\ds{\sum_{\ss N_2=2}^{l}}}{\ds{\sum_{\ss\{N_2\}}}}{\ds{\sum_{\ss{\rm words}}}}
\px A(\xi,t)\o (N-l+1)2^{l-2}}\no$$
where $l\geq2$. The first sum is over the possible beginning point, $i$, 
of the schemata and the following two sums represent the different 
configurations of any number $N_2 \leq l-2$ specified bits chosen among the $l-2$
available sites. The number of available sites is $l-2$ because we fix the 
ends bits.

Using (\docref{maseqtwo}) one may derive a recursion relation for the
expectation value of the observable $A$, from
$$<A(t+1)>_l= {{\ds{\sum_{i=1}^{\ss (N-l+1)}}}
{\ds{\sum_{\ss N_2=2}^{l}}}{\ds{\sum_{\ss\{N_2\}}}}
{\ds{\sum_{\ss{\rm words}}}}P(\xi, t+1) A(\xi,t+1)\o (N-l+1)2^{l-2}}\no$$
We now encounter a difficulty: time dependence enters in the above 
equation not only in the changing probability distribution $P(\xi, t+1)$,
which can be substituted using (\docref{maseqtwo}), but also in $A(\xi,t+1)$. 
This occurs even though many observables of interest are time-independent
functions of the string states as the summing over degrees of freedom 
associated with passing to a more coarse grained description induces an
implicit time dependence in the coarse grained observables. For example 
$\fxi$ is a population-dependent observable even though $f(C_i)$ is not.

One can derive a function on schematas, such as
schemata fitness, via a population average with the string probability 
$P(\C_i, t+1)$, for which one has the microscopic evolution equation, but 
this clearly leads to a very complicated calculation. 
To simplify matters, to search for structure in the population 
we define a time-independent function on schematas. The particular function
we choose is the average selective advantage that in-schemata 
bits would enjoy if the schemata were immersed in a random population, 
\eqlabel{\deltafxi}
$$\delta f_{\xi} = \left( {N \over N_2}\right) \left( {1 \over 2^{\ss N-N_2}}
\right) \sum_{\ss \eta -{\rm words}}^{2^{\ss N-N_2}}
\left( f(\xi, \eta) - {1 \over 2} \right),\no$$
where $\eta$ represents the out-of-schemata bits and the average 
fitness in a random population has been normalized to $1/2$. 
Note that here, and in the rest of the paper, we are looking at the fitness 
deviation per schemata bit
as opposed to section 3 where the total fitness of a schemata was being
considered. This observable corresponds to the {\it effective fitness}
of in-schemata bits either if the population {\it is} in fact random, or
if the landscape assigns an independent fitness contribution to each bit 
in the chromosome ($k = 0$ in the terminology of the
Kauffman $Nk$-model). In general, it is 
a useful test function with which one can probe for the emergence of structure during 
the first steps of evolution away from a random initial population. 
We will refer to this observable below as the {\it in-schemata fitness}.

We will make use below of the following simplified averages: if $A(\xi)$ is 
independent of the initial defining point of the schemata, 
or if the landscape is ``generic'', then we can sum over this point to find
$$<A(t)>_l={1\o 2^{l-2}}   {\sum_{N_2=2}^{l}\sum_{\{N_2\}}\sum_{\ss{\rm words}}
\px A(\xi) }\no$$
By ``generic'' we mean that within the class of landscapes we are considering, 
such as an $Nk$-model for a particular value of $k$, there is no systematic
bias in the fitness function for a particular part of the string, i.e. 
the sums over words, configurations and $N_2$ leads to an average which 
is effectively translation invariant, i.e. the system is effectively self-averaging. 
Similarly, if $A(\xi)$ depends only on the order of the schemata, $N_2$, one has
$$<A(t)>_l={1\o 2^{l-2}}  {\sum_{N_2=2}^{l}{}^{l-2}C_{N_2-2}\px A(\xi) }\no$$
We will also use the notation $<<A>>_l$ to represent the 
average over schematas and over crossover points, namely
$$<< A >>_l = {1 \over l-1} \sum_{i=1}^{l-1} <A>_l\no$$

Considering the expectation value of the 
in-schemata fitness, the equation which gives the improvement of 
$<\delta f_{\xi}>_l$ from generation $t$ to generation $t+1$ is 
$$\Delta_l=<<\dfx{\delef\o\fav}>>_l\no$$
where $\delef=\ef-\fav$. More explicitly, using the evolution equation for schematas,
one finds
$$\eqlabel{\deltaev}
\Delta_l=<{\dfx\dfxi\o \fav}>_l-p_c\left({l-1\o N-1}\right)<<{\dfx\o P(\xi,t)}
\left(\pxf-\pxLf\pxRf\right)>>_l,\no$$
where $\dfxi=\fxi-\fav$. The first term is independent of $l$ in a random
population if the fitness landscape itself is $l$ independent. All the 
$l$ dependence lies in the crossover terms. If the 
parts of a schemata are selected more than the whole it is clear that the net 
contribution from crossover will be positive.
 
It worth pausing here to consider the meaning of the quantity
$\Delta_l$. As we defined it, $\Delta_l$ measures the average improvement
of the in-schemata fitness over one step of evolution. How does this
improvement come about? First of all, schematas with $\dfxi> 0$
will be more frequent in the parent population, thanks to the selection
factor $(1 + s)$, where
$$s = {\dfxi\o \fav} ={2N_2 \over N}\dfx. \no$$
where the latter equality is only true for a random initial population or 
$k=0$ model. The next step is to consider the action of the crossover 
operator. On the one hand selected parents with $\xi$ may not pass it on to 
their offspring if crossover ``breaks'' the schemata. However, 
there is a possibility that $\xi$ be reconstructed from parents that 
have parts of $\xi$ but not all of it. This reconstruction term gives a 
positive contribution because if $\xi$ has a selective advantage then subsets
of $\xi$ will, for an average type of landscape, be more likely than not to 
have some selective advantage as well, so the parts of $\xi$ that are 
needed for reconstruction are available in the parent population with 
an enhanced probability. The key question is, of the destruction and the 
reconstruction terms which is larger for a particular value of $l$? Before 
we turn to answering this question in particular cases, let us consider the
relation between $\Delta_l$ and the spatial correlation function.

If the population is uncorrelated, in other words if $P(\xi,t) = \prod_i 
P(\xi_i,t)$, where $\xi_i$ is the $i$'th bit of $\xi$, 
then the expectation value of $\dfx$ is independent of $l$, as 
$N_2 \dfx = \sum_i \delta f_{\xi_i}$ and
$$<\sum_i \delta f_{\xi_i}>_l\no$$ 
is just the uncorrelated sum of  contributions from 1-schematas.
The fact that the existence of correlations in $P(\xi, t+1)$ implies an 
$l$ dependence can be demonstrated explicitly. One writes
$$\ef \approx {1 \over N_2} \sum_{i=1}^{N_2} f_1(\xi_{i})
+ {1 \over N_2(N_2-1)} \sum_{i=1}^{N_2} \sum_{j \neq i} \left( 
f_2(\xi_{i} \xi_{j}) - {1 \over 2}(f_1(\xi_{i}) + f_1(\xi_{j})) \right), \no$$
where 
$$f_1(\xi_i) = {1 \over 2^{N_2-1}} \sum_{\{\xi_k, \ k \neq i\} } \ef(\{\xi_k\}) \no$$
$$f_2(\xi_i \xi_j) = {1 \over {2^{N_2-2}}} \sum_{ \{\xi_k, \ k 
\neq i,j \} } \ef(\{\xi_k\})\no$$
and we are considering only up to two-point correlations.
Defining $\delta s_{\ss\xi}= \delef-\dfx$, which is a measure of the 
selective advantage over and above the in-schemata fitness, one finds
$$\delta s_{\ss\xi} \approx {1 \over N_2} \sum_{i=1}^{N_2} f_1(\xi_{i})
+ {1 \over N_2(N_2-1)} \sum_{i=1}^{N_2} \sum_{j \neq i} \left( 
f_2(\xi_{i} \xi_{j}) - {1 \over 2}
(f_1(\xi_{i}) + f_1(\xi_{j})) \right)-\fav-\dfx.\no$$
For a $k=0$ landscape, 
$$\delta s_{\ss\xi}\approx {1 \over N_2(N_2-1)} \sum_{i=1}^{N_2} \sum_{j \neq i} \left( 
f_2(\xi_{i} \xi_{j}) - {1 \over 2}(f_1(\xi_{i}) + f_1(\xi_{j})) \right)\no$$
So, in this case we see that the existence of a selective advantage 
is due to the existence of correlations in the effective fitness. 
Defining a selective coefficient $s_l$ which represents the selective 
advantage for a schemata to be of size $l$ one finds
$$\Delta_l = <<\delta f_{\xi}^2 >>_l (1 + s_{\ss l})\no$$
where
$$s_{\ss l}={<<\dfx^2\delta s_{\ss\xi}>>_l\o <<\dfx^2>>_l}.\no$$
In this expression for $\Delta_l$, $<<\dfx^2>>$ is independent of $l$ for
a random initial population. Thus we see that any $l$ dependence 
can be attributed to the existence of spatial correlations. 
 
If the reconstruction term from crossover exceeds the 
destruction term for some $l$, then from the above 
one concludes that the fitness 
improvement attributed to a particular bit in the string depends 
on it being part of selected schematas of this size. The maximum
value of $\Delta_l$ is attained when the contribution of an individual bit 
is most enhanced by the fact that this bit belongs to strings 
that include other specified bits at a distance at most equal to $l$,
namely those strings which include a selected schemata of size $l$. 
That the conditioning information on the existence of other specified
bits should be useful is a direct consequence of the correlations
between the different bits in the string. 
The reason why we emphasize the relation between $\Delta_l$ and the 
correlation function is that correlations are intimately linked to the 
emergence of effective degrees of freedom. In this sense, the function 
$\Delta_l$ is related to the expected size distribution of the effective 
degrees of freedom. 

\vskip 0.3truein

\line{\bgg 5. Asymptotic Solutions \hfil}

In this section we consider some asymptotic solutions of the evolution equation for 
$\Delta_l$ derived in section 4. In particular we will consider two limiting cases: 
the evolution of schematas starting from a completely random initial state; 
and a random perturbation around a completely ordered state. As one of our 
principal considerations is in investigating the validity of the building block 
hypothesis we will set the mutation rate to zero as the 
effects of the latter do not depend on schemata size. We will derive expressions
for $\Delta_l(t+1)$ and $\Delta_l(t+2)$ starting out from an initial random population at 
time $t$. 

For a random initial population at time $t$,
$$\Delta_l(t+2)=<<\dfx(t+2)>>_l-\Delta_l(t+1)\no$$
Even though $\dfx$ is time independent we use the above notation to indicate
that its expectation value is with respect to the probability distribution at time
$t+2$.

In the initial random population, the effective schemata fitness is
the in-schemata fitness $\delta f_{\xi}$ and 
$$\dfxi = {N_2 \over N} \delta f_{\xi},\no$$
$$P(\xi, t) = {1 \over 2^{N_2}}.\no$$
Thus one finds
\eqlabel{\deltaone}
$$\eqalign{\Delta_l(t+1)&=\left(1-p_c({l-1\o N-1})\right)<<\alpha >>_l +\cr
&p_c({l-1\o N-1})<<{N_L\o N_2}\beta_L+{N_R\o N_2}\beta_R+{4N_LN_R\o N^2}
\dfx\dfx_L\dfx_R>>_l\cr}\no$$
where we have introduced the notation for the quadratic terms
$$\alpha =  {1 \over 2^{N_2}} \sum_{\s\rm words} {2 N_2 \over N} 
\delta f_{\xi}^2 , \no$$
$$\beta_L = {1 \over 2^{N_2}} \sum_{\s\rm words} {2 N_2 \over N} 
\delta f_{\xi} \delta f_{\xi_L}, \no$$
with an analogous expression for $\beta_R$.

 As one of our principle purposes here is to examine the block hypothesis in light
of the evolution equation we have derived, and the associated notions of coarse
graining and effective degrees of freedom, we will try to derive explicit results in some
concrete cases based on generic fitness landscapes. The Kauffman $Nk$-models provide
such a set of landscapes. Here we will specialize to the case $k=0$ which
is neutral in the sense that it neither favours nor disfavours correlations 
between bits. We will discuss how our results generalize to a $k=2$ landscape
in the next section. In the $k=0$ landscape,
$$\delta f_{\xi} = {N_L \over N_2} \delta f_{\xi_L} + 
{N_R \over N_2} \delta f_{\xi_R}. \no$$
We also have that $<<\dfx\dfx_L\dfx_R>>_l=0$ which results in the complete cancellation
of the destruction and reconstruction crossover terms the final result being
$$\Delta_l(t+1)=<<\alpha>>_l.\no$$

The above expression is for an arbitrary $k=0$ landscape. In order to find 
a more explicit solution we must consider a more explicit landscape. 
We will consider two: a binary landscape where the fitness of a bit 
may only take two values, $1$ and $0$; and a landscape where the 
fitness of a bit is selected uniformly at random from the interval 
$[0,1]$. Both landscapes conform with the requirement that the average 
fitness per bit in a random population is $1/2$. Let $x_i$ 
denote the deviation 
from the mean fitness of bit number $i$, i.e. $x_i =  f_i - 1/2$. We find 
$$N_2 \dfx = \sum_{i=1}^{N_2} (2\xi_{n_i}-1) \ x_{n_i},$$
where the indices $n_i$ denote the specified bits of the schemata 
($i = 1, \cdots, N_2$). Squaring this expression and using 
$\sum_{i\neq j}<(2\xi_{n_i}-1) (2\xi_{n_j}-1)> = 0$ one finds
$$<N_2 \dfx^2> = <{1 \o N_2} \sum_{i=1}^{N_2} x_{n_i}^2>.$$
The averaging over configurations then gives, for $l \geq 3$, 
$$<<\alpha>>_l = {2 \o N(N-l+1)l(l-1)(l-2)}\ \sum_{i=1}^N \ m_i\ x_i^2,$$
where
$$m_i = l(l-1)(l-2) \ \ \ \ \ \ \ \ \ \ \ \ \ \ (l \leq i \leq N-l+1),$$
$$m_i = (l^2-3l)+i(l^2-5l+8)+{l-2i \o 2^{l-2}} \ \ \ \ \ (i < l),$$
and symmetrically for $i > N-l$. 

 For the case of a binary landscape the final answer is
$$<<\alpha>>_l={1\o 2N}.\no$$
In the random landscape for large $N$ we can assume that
the average over the $N$ bits (weighted by $n_i$) can be replaced by an
average over the distribution of $x_i$ used to generate the landscape. Then,
$$<<\alpha>>_l={1\o 6N}.\no$$
Thus one sees that crossover acts in a scale invariant way at the first
time step of evolution from a random initial population: there is no 
preference whatsoever for small blocks at the expense of large blocks. 

We will now consider what happens at time $t+2$. The extra ingredient we need 
relative to the above calculation is $<<\dfx(t+2)>>_l$. To calculate this 
we in turn need to calculate $P'(\xi,t+1)$, $P'(\xi_L,t+1)$ and $P'(\xi_R,t+1)$ 
i.e. the selection probabilities at time $t+1$, calculation of which requires 
knowledge of ${\bar f}(\xi,t+1)$, ${\bar f}(\xi_L,t+1)$ ${\bar f}(\xi_R,t+1)$ 
and ${\bar f}(t+1)$. Specializing once again to a $k=0$ landscape,
one finds
$$P(\xi,t+1)={1\o 2^{N_2}}\left(1+{2N_2\o N}\dfx+{p_c\o N-1}\sum_{k=1}^{N-1}
{4N_LN_R\o N^2}\dfx_L\dfx_R\right),\no$$
$${\bar f}(t+1)=(1+2\alpha_s){\bar f}(t)\no$$
and
\eqlabel{\peeprime}
$$\eqalign{&P'(\xi,t+1)={1\o 2^{N_2}(1+2\alpha_s)}\left[(1+{2N_2\o N}\dfx)^2+
2{(N-N_2)\o N}\alpha_{\ss (N-N_2)}\right.\cr
&\ \ \ \ \ \ \ \ \ \ \ \ \ \ \ +{p_c\o N-1}(1+{2N_2\o N}\dfx)\sum_{k=1}^{l-1}
{4N_LN_R\o N^2}\dfx_L\dfx_R\cr
&\left.\ \ \ +{4p_c\o N(N-1)}\left(\sum_{k=1}^{l-1}(N_R\dfx_R\beta_{\ss(k-N_L)}+N_L\dfx_L
\beta_{\ss (N-k-N_R)}+N_2\dfx(\sum_{k<\xi}\beta_k+\sum_{k>\xi}
\beta_{\ss (N-k)})\right)\right]
\cr}\no$$
where
$$\alpha_{\ss (N-N_2)} =  {1 \over 2^{N-N_2}} \sum_{\s \eta-{\rm words}} {2 (N-N_2 )\over N} 
\delta f_{\eta}^2 \no$$
\eqlabel{\toleft}
$$\beta_{\ss(k-N_L)}={1 \over 2^{k-N_L}}\left({2 (k-N_L )\over N}\right)^2 
\sum_{\s \eta_L-{\rm words}}\delta f_{\eta_L}^2. \no$$
\eqlabel{\lessthan}
$$\beta_k={1 \over 2^{k}}\left({2 k\over N}\right)^2
\sum_{\s L-{\rm words}}\delta f_{L}^2. \no$$
In (\docref{lessthan}) the sum is over words associated with bits to the left of the crossover
point given that all the schemata $\xi$ lies to the right of the crossover point. The expression
for $\beta_{\ss (N-k)}$ is analogous but with the sum over words being associated with bits
to the right of the crossover point given that the schemata 
lies to the left. Equation 
(\docref{toleft}) is associated with a sum over words for the bits to the left of the crossover
point but excluding bits that are in the schemata. Likewise the expression  
$\beta_{\ss (N-k-N_R)}$ contains a sum over words associated with bits that are out of
schemata to the right of the crossover point. Finally, 
$\alpha_s=1+(2N_2/N)\alpha+(2(N-N_2)/N)\alpha_{\ss (N-N_2)}$.

If one considers $\xi_L$ and $\xi_R$ as schematas on exactly the same footing as $\xi$
then the expressions for $P'(\xi_L,t+1)$ and $P'(\xi_R,t+1)$ are completely analogous to 
those above except that one is now considering the bits of $\xi_L$ and $\xi_R$ which lie
to the left and the right of the crossover point. Combining these expressions with 
equations (\docref{peeprime}-\docref{lessthan}) after some lengthy but
straightforward calculations one finds
\eqlabel{\deltafin}
$$\eqalign{\Delta_l(t+2)=&\left({1-2\alpha_s\o 1+2\alpha_s}\right)<<\alpha>>_l
+{p_c\o (1+2\alpha_s)(N-1)}<<{2N\o N_2}(\beta_R\beta_k+\beta_L\beta_{\ss (N-k)})>>_l\cr
&+{2p_c\o (1+2\alpha_s)(N-1)}<\alpha\sum_{k<\xi}\beta_k+
\alpha\sum_{k>\xi}\beta_{\ss(N-k)}>_l\cr}.\no$$
The first term in the right hand side of (\docref{deltafin}) 
is the result of selection at $t+1$
on the population that was the result of selection at $t$. It is crossover independent as
is manifest in the fact that $p_c$ does not appear. The last two terms are associated 
with the effects of selection on the population at time $t+1$ which has 
incorporated non-trivial contributions from crossover at time $t$. More precisely,
the picture is the following: $k=0$ selection on a random population induces
anti-correlations in $P'(\xi_L, \xi_R, t)$ when both $\dfxL$ and $\dfxR$ are
positive due to the quadratic term $\sim \dfxR \dfxL$. Crossover reduces these
anticorrelations, thereby enhancing the {\it whole} schemata $\xi = \xi_L + \xi_R$
relative to its parts. Selection at $t+1$ reinforces this effect of crossover 
to enhance $\xi$, leading to the net positive contribution  to $\Delta_l(t=2)$.

As above, we will consider the binary landscape and a landscape where the 
fitness of a bit is selected uniformly at random from the interval 
$[0,1]$. Similar calculations to the ones given above for $<<\alpha>>_l$
lead to the following expressions for 
$<<{2N\o N_2}(\beta_R\beta_k+\beta_L\beta_{\ss (N-k)})>>_l$
and $<(\alpha\sum_{k<\xi}\beta_k+\alpha\sum_{k>\xi}\beta_{\ss(N-k)}>_l$, 
for $l \geq 4$:
$$<<{2N\o N_2}(\beta_R\beta_k+\beta_L\beta_{\ss (N-k)})>>_l={c\o 108N^3}(N-l)(N-l-2)\no$$
$$<(\alpha\sum_{k<\xi}\beta_k+\alpha\sum_{k>\xi}\beta_{\ss(N-k)}>_l=
{c(l-1)\o 72N^3}-{c(l^2-5l+8)\o 216N^3}+{c\o 27N^42^l}\no$$
where $c= 3$ for the binary landscape and $c=1$ for the random landscape.
Putting these terms together we arrive at the final expression for 
$\Delta(t+2)$,
\eqlabel{\delfinfin}
$$\Delta(t+2)=\left({3N-c\o 3N+c}\right){c\o 6N}+
p_c\left({2N^2-Nl+l^2+N+l-8-(8/2^l)\o 144(3N+c)N^2(N-1)}\right)\no$$
>From this expression one can readily see that the effects of crossover 
are always positive, i.e. the effects of schemata reconstruction 
outweigh those of schemata 
destruction. A graph of $\Delta(t+2)$ versus $l$ can be seen in figure 1.

We now turn our attention briefly to the limiting case of an almost 
organised population. In this limit, one can consider that the strings 
differ from the population-consensus at most at one site; we will refer
to the differing site as a ``defect''. There are $N$ possible 
defects each with an effective negative fitness differential over the
consensus string. The evolution equation implies an
equation for the growth or decay of defects, where one 
immediately sees that the effect of crossover is strictly neutral:
there is no net creation or destruction of defects by pure crossover
without selection. In other words the geometrical effect of crossover is
zero. The role of crossover in this limit is only to mix the defects in the 
population. So in this limit $\Delta_l$ is once again strictly independent
of $l$. The possibility of multiple defects in a single string 
raises the possibility of correlations in the distribution of
defects along the string, which would induce mirroring  
correlations in the schematas, so $\Delta_l$ may acquire a non-trivial
$l$-dependence as a second order effect in the mean density of
defects, which is the perturbative expansion parameter near the 
ordered limit. 

 Taking once again the $k=0$ landscape, the fitness penalty per bit for 
two defects is given by $2\delta f_{ij} = \delta f_i + \delta f_j$, so
$${P'(ij, t) \o P'(i ,t) P'(j, t)} \sim {{1 + {2 \delta f_{ij} \o {\bar f}N}}
\o {(1+{\delta f_i \o {\bar f}N})(1+{\delta f_j \o {\bar f}N})}}
= \left(1+{\delta f_i \delta f_j \o {\bar f}^2 N^2} \right)^{-1}.$$
Since $\delta f_i \delta f_j > 0$, $P'(ij, t) < P'(i, t)P'(j,t)$: 
selection induces an anticorrelation between the defects. 
Now, crossover 
enhances $P(ij, t+1)$ to bring it closer to $P(i, t+1)P(j, t+1)$. Since
the schemata ($ij$) is more strongly damped than $i$ or $j$ separately,
the selection at the next time step will destroy more defects than without
crossover. So here again as near the random limit  crossover has a 
beneficial effect due to the enhancement of whole schematas relative to 
its parts. Near the random limit this was beneficial because the whole 
schemata was picked up by selection, here it
is beneficial because the whole schemata is more strongly damped by 
negative selection so defects die out more rapidly.

 One can think of the initial random population as being
the high temperature fixed point of the model, given that every point 
in configuration space is equally occupied; in this limit the correlation
length is zero. The ordered limit would then naturally be interpreted as 
the low temperature limit. Our results from this section can be summarised 
by saying that crossover is a net positive contributor to fitness growth
at second order near both temperature limits, $T \to 0$ and $T \to \infty$. 

\vskip 0.3truein

\line{\bgg 6. Effective degrees of freedom in the $N2$ landscape \hfil}

 The $k=0$ landscape discussed in the last section has the virtue of being
``neutral'' from the point of view of the block hypothesis, however it 
is not a realistic example of landscapes usually encountered in complex 
optimization problems: It is strongly correlated, has a single optimum 
and does not present frustration we will therefore turn our attention now
to Kauffman's $Nk$ model with $k=2$.
There are two mechanisms by which connected landscapes 
can induce correlations: 
On the one hand, schematas that contain landscape-related bits 
have a sharper selective coefficient because there are fewer unspecified 
bits involved in their fitness contribution. On the other hand, the balance 
between the schemata destruction and reconstruction terms from 
crossover is broken to first order. For example if the effective fitness 
of a whole schemata is less than the sum of the effective fitnesses of 
its parts, the growth of a schemata can be magnified by breaking it down 
into parts, growing the parts and then reconstructing the schemata. 
We will analyse both of these correlating effects below.

 The $Nk$ model can be described as follows: the fitness function is
specified by giving, for each bit of the string, $k$ connections
to as many other bits and a table of $2^{k+1}$ 
random numbers uniformly distributed in the unit interval. To compute the 
fitness contribution of one bit of a string one forms a $(k+1)$-word 
from this bit and its connected partners and translates it into an 
integer $n \in \{1, 2, \cdots, 2^{k+1}\}$. The fitness is then given by 
the entry number $n$ in the table of random numbers. Note that the $k$
connections and the table of random numbers are chosen independently for each
bit in the string. For $k=0$ the fitness contribution from each bit is 
independent of the others and takes one of two possible values.
The resulting landscape gives rise to a unique extremum 
(barring accidental degeneracies). At the other extreme, $k=N$, 
the fitness of every string is an independent random number (random 
landscape). The ruggedness of the landscape increases with $k$, which 
allows one to use $k$ as a free parameter in order to be able to 
model real landscapes of arbitrary ruggedness. Here we will consider 
only the representative case $k=2$. 

 Let us first restrict our attention to schematas of two definite
bits ($N_2 = 2$).  
There are three possible situations for a 2-schemata. Either the two bits are 
not connected by the fitness landscape, one bit is the connected 
partner of the other, or the bits are connected both ways. This last situation
is improbable for $N >> 1$, so we will focus on the first two cases. If 
one has two unrelated bits in an otherwise random initial population, the 
effective fitness of each bit in this schemata is equal to an average of four 
of the eight random numbers in the fitness table at that site, because one
of the three bits is fixed and the other two are picked at random. If, on the
other hand, one of the bits is connected to the other then its fitness
contribution is given by averaging over the two possible values that the 
other connected partner can take. This is an average
of two out of the eight random numbers in the fitness table. 
The key point is that the average of two random numbers typically differs
from ${1/ 2}$ more than the average of four. Therefore,
schematas which include landscape-related bits will have a stronger
selective coefficient, in absolute value. This leads to a bias for the 
condensation of schematas that recognize the structure of the fitness 
landscape. 

In order to make this argument more precise, we need to compute the 
expectation value of the best of $m_1$ averages of $m_2$ random numbers,
where each random number is uniformly distributed in the unit interval.
The probability distribution of the best of $m_1$ averages of $m_2$ 
random numbers is equal to the derivative of the probability that
$z$ is larger than all $m_1$ averages. If we call $P(x_1, \cdots, x_{m_1})$
the distribution of the averages, the probability that $z$ is greater than
all of the averages is
$$p(z > sup(m_i)) = \int_0^1 dx_1 \cdots \int_0^1 dx_{m_1} \ P(x_1, \cdots, x_{m_1})
\ \prod_{i=1}^{m_1} \theta(z - x_i) .\no$$
Since the ${m_1}$ averages are statistically independent in this case, 
this expression reduces to
$$p(z) = \left( \int_0^1 dx P(x) \theta(z - x) \right)^{m_1}.\no$$
The expectation value of the best of the $m_1$ averages is 
$$<z_{\ss\rm max}> = \int \ z p'(z) dz\no$$

For our purposes it is sufficient to consider the cases where $m_1, 
m_2 \in \{ 2, 4\}$. 
For $m_2 = 2$ the distribution of the average of two uniformly 
distributed random numbers is given by
$$P(x) = 4x \ \ \ \ \ {\rm for}\ \ \ \ \ x < 1/2$$
and the symmetry condition $P(1/2+x) = P(1/2-x)$. The 
expectation values for the best of $m_1$ such averages are:
for $m_1 = 2$, $m_2 = 2$, \ $<z_{\ss \rm max}> = 0.6167$;
while for $m_1 = 4$, $m_2 = 2$, \ $<z_{\ss\rm max}> = 0.7300$.
For $m_2 = 4$ (averages of four uniformly distributed variables),
one has
$$P(x) = {128 \over 3} x^3 \ \ \ \ \ {\rm for}\ \ \ \ \ x < 1/4,\no$$
$$P(x) = {128 \over 3} x^3 - {2 \over 3} \ \ \ \ \ {\rm for}\ \ \ \ \ 1/4 < x < 1/2,\no$$
and the symmetry condition given above. One finds the following result:
for $m_1 = 2$, $m_2 = 4$, \ $<z_{\ss\rm max}> = 0.5673$.
Finally, the expectation value of the best of $m_1$ uniformly 
distributed random variables is
$$<z_{\ss\rm max}> = {m_1 \over m_1+1}.\no$$ 
Here we will need only the best of eight, which is equal
to $8/9 = 0.8889$.

As mentioned previously, we will consider only schematas
with two definite bits. If the two bits are not related by a landscape
connection the effective fitness of any one of these bits in a random 
population is given by the average of four random numbers from the
fitness table, where the averaging is over the values of the two connected
partners which determine the fitness contribution of this bit. Thus, 
the best schemata can be expected to have a selective advantage
$$s_1 = {{\bar f}_{\xi} \over {\bar f}} - 1 = {4 \over N} (0.567 - 0.5).\no$$
Now, if there is a landscape connection between the two bits
of the schemata, the contribution of one of these bits to the string
fitness is given by an average of two random numbers, since we only
need to average over the other connected partner which is not in the
schemata. The best schemata in this case will have a selective 
advantage 
$$s_2 = {4 \over N} \left({0.567 + 0.73 \over 2} - 0.5\right).\no$$

In the case $N=40$ analyzed in the previous section, the ratio of 
the growth rates of a 2-schemata which recognizes a landscape connection 
to that of one that doesn't is
$$r = {1+s_2 \over 1+s_1} = 1.0081.\no$$
This result should be compared to the effect of crossover which we 
computed in the $k=0$ landscape at the second time step: $1+s_l$ was found
to fluctuate between $1.0025$ at $l=N/2$ and $1.0029$ at $l=2$, $l=N-1$. 
Clearly, the conclusion is that landscape correlations should be taken 
into account in a proper analysis of the condensation of ``schematas''.

In our discussion we neglected the possible existence of frustration
and assumed that the fitness contribution of the two bits of the
schemata could be optimized independently without affecting the
mean fitness contribution of the other bits in the string. A
more careful analysis including frustration would be much more
complicated; however one expects that at least for small $N_2$
frustration should be marginal and that our conclusions should 
hold qualitatively. Of course, there are fewer 2-schematas that recognize a 
landscape connection than not, so the overall contribution of such schematas
to the condensation of effective degrees of freedom is diluted by 
a phase space factor $2/N$, relative to 2-schematas of landscape-related
bits. Thus, one expects that the first stage of divergence from a random 
population will be dominated by schematas which do not understand the fitness
landscape. The landscape-related schematas, which grow at
a faster rate, will eventually overcome the contrary phase 
space factor and become more important in the condensation process. 

 Returning to the fundamental equation (\docref{maseqone}) for the growth 
of in-schemata fitness we can evaluate the effect of crossover in a $k =2$
landscape by calculating $\Delta_l$ in the first step away from a 
random population:
$$\Delta_{\ss N_2=2}^{(n)} = < {4 \o N} \dfx^2>^{(n)} - {4p_c \o N(N-1)} <l-1>^{(n)}
<\dfx\left( \dfx - {N_L \o N_2}\dfxL - {N_R \o N_2} \dfxR \right)>^{(n)},$$
where we have used the identity, valid in a random population, 
$\sum_{\ss words} \dfxL \dfxR\dfx = 0$,
and the average $<>^{(n)}$ runs over the set of all schematas with $N_2 = 2$ 
definite bits with $n=0,1,2$ landscape connections between schemata bits. 
We are also assuming that there is no explicit $l$ dependence in the fitness
landscape itself.

 The evaluation of $\dfx$ depends on the 
number of in-schemata connections. One must evaluate the contribution of 
each of the two bits in the schemata. If there are no in-schemata connections 
then the averaging over unspecified bits leads to a 
contribution to $\dfx$ equal to the average of four 
of the eight random numbers in the fitness table. If one of the 
bits is connected to the other, then in evaluating its fitness
contribution one has only one unspecified bit and the contribution to $\dfx$ 
turns out to be the average of two of the eight random 
numbers. The values of $\dfxL$ and $\dfxR$ are always given by the
average of four random numbers. 

 Thus, if there are no in-schemata
connections then $\dfx = {N_L \o N_2}\dfxL + {N_R \o N_2} \dfxR$ 
and the contribution of the crossover term vanishes as in the $k=0$ case.
If we denote by  $\sigma^2$ the variance of the random number distribution
used to generate the tables of eight possible fitness contributions 
for each bit, the averaging over schematas with $n=0$ landscape connections
gives
$$\Delta_{\ss N_2=2}^{(0)} = <{4 \o N} \dfx^2>^{(0)} = {\sigma^2 \o 2N}.$$
On the other hand, if there is one in-schemata connection
then $<\dfx^2>^{(0)}$ is the variance of the average of two random numbers
plus the variance of an average of four, while one of the terms 
$<\dfx \dfxL>^{(0)}$ or $<\dfx \dfxR>^{(0)}$ is equal to half the variance of 
two random numbers, the other being the variance of an average of four.
Using $<l-1> = (N+1)/3$, one finds
$$\eqalign{
\Delta_{\ss N_2=2}^{(1)} &=  
\left({3 \o 4} -{p_c (N+1) \o 12(N-1)}\right) {\sigma^2\o N} \cr
 &= \Delta_{\ss N_2=2}^{(0)} + \Delta_{\ss N_2=2}^{(0)} \left(
{1\o 2} -{p_c (N+1) \o 6(N-1)}\right) .\cr}$$
Similarly, for $n=2$ in-schemata connections,
$$\Delta_{\ss N_2=2}^{(2)} - \Delta_{\ss N_2=2}^{(0)} =  \Delta_{\ss N_2=2}^{(0)} 
\left(1 -{p_c (N+1) \o 3(N-1)}\right).$$
In these expressions, the $p_c$-independent correction is the result of the 
selective advantage of schematas that recognise landscape 
connections, which we discussed previously. These numbers appear somewhat
magnified relative to $r$. This is only because here we are examining the
in schemata fitness per bit whereas $r$ was associted with the growth rate
of the entire string.
The crossover contribution reduces this correlating
effect of the landscape but only by a factor of $2/3$ in 
the limit $p_c \to 1$, $N \to \infty$. 
In conclusion, schematas which reflect the landscape connections 
contribute more (per bit) to the growth of fitness than schematas involving 
unrelated bits. 

 A similar conclusion can be expected to hold if one considers larger 
schematas with $N_2 > 2$. Extending the argument to general
schematas one is led to consider {\it fitness trees}: the fitness
tree of a bit is the set which consists of the bit
itself, its connected partners, the connected partners of the 
these connected partners, and so on. We can define an 
order $n$ truncated fitness tree by truncating this procedure after $n$ 
steps. The dominant value of $n$ depends on the degree of order in the 
system, which is a function of the mutation rate. For a high mutation
rate one expects the gene pool to be highly disordered and 
effective degrees of freedom are mostly single bits ($n=0$) 
or truncated fitness trees with small values of $n$. As the 
mutation rate decreases larger trees can condense and the dominant 
value of $n$ increases. 
This leads us to propose the following conjecture 
on the nature of the effective degrees of freedom, which we shall 
call the ``fitness tree hypothesis''.

\item{} $\bullet$ {\it The effective degrees of freedom 
of genetic algorithms with $Nk$ fitness landscapes are the truncated 
order $n$ \ fitness trees. The effective value of $n$ increases as the 
condensation process allows for an increasingly structured gene pool}.


 In order to test this hypothesis we designed a numerical simulation
with a population of 1000 individuals in an $Nk$ landscape, with
$N=40$ and $k=2$. The crossover probability was taken to be equal to one. 
The spatial correlation function 
measures the correlation of bits at distance $d$ along the string and 
tests the block hypothesis directly. A second correlation function
measures this correlation as a function of the connective distance 
between bits, defined as the smallest number of landscape connections
from one bit to the other. The results are shown in [Figures 2 a-c]. 
At generation 15 ([Fig. 2a]) the spatial correlation function reflects the
preference for small schematas, as suggested by the block hypothesis.
After 100 generations ([Fig. 2b]) the spatial correlation function has become 
weak and roughly independent of the distance; on the other hand the correlation
of landscape-related bits becomes significant at connective distance one.
By generation number 150 one finds statistically significant correlations 
up to connective distance four, which are progressively reinforced. 
In [Figure 2c] we show the correlation functions at generation 200. 
Since the mutation rate is equal to zero in these simulations, population 
diversity eventually decreases and becomes insufficient to derive 
statistically relevant correlation coefficients. At generation 350 
the strings are totally condensed up to connective distance two 
(the first two correlation coefficients are equal to one); the gene 
pool is completely organized at the 500'th generation. 

Throughout this article, with the exception of the numerical experiments 
finite size effects were neglected. If one considers their contribution 
the failure of the block hypothesis only becomes more apparent. Here we will 
mention only briefly two arguments to this effect.

 In a finite population the difficulty of {\it finding} a good schemata 
must be considered, since not all schematas are present in the 
initial population. Since the number of schematas with fixed $N_2$ grows
with $l$ as $^{l-2}C_{N_2-2}$, one expects it to be easier to discover good 
large schematas than small ones. Another important
finite size effect is the effective non-linearity of selection emphasised
in the Neutral Theory of Molecular Evolution \citelabel{\kimura}[\cite{kimura}]: 
Schematas with only weak
selective coefficients are not necessarily selected, as the neutral drift 
due to fluctuations in the selection of parents dominates over selection
unless $| s_{\ss eff} | > 1/P$, $P$ being the effective breeding population. This 
leads to an effective non-linearity of selection due to the existence of a 
threshold in favour of schematas with
a selective coefficient above this value. Since the selective
coefficient of a schemata grows in proportion to $N_2$, this effect favours
schematas with large $N_2$. Combining this result with the previous comment
on the probability to find good schematas being proportional to 
$^{l-2}C_{N_2-2}$, we find that schematas with small values of $l$ are 
strongly disfavoured by the finite size effects.

\vskip 0.3truein
\vfill\eject

\line{\bgg 7. Conclusions \hfil}

The bulk of this paper has been devoted to deriving 
equations that describe the evolution of string populations 
in genetic algorithms, and in particular how effective
degrees of freedom may emerge during this evolution.
We started with an equation that governed the evolution of the
strings themselves under the joint action of selection, mutation
and crossover. We found that this equation could be elegantly
expressed in terms of the evolution of a string $C_i$ and its
subcomponents relative to the crossover point, $C_i^L$ and $C_i^R$. 
This naturally introduced the notion of a coarse graining
relative to a description in terms of the strings themselves, the 
coarse graining being associated with sums overs strings that
contained a part of $C_i$.
Subsequently we derived an analogous equation for the evolution of
schematas, this time in terms of a schemata and its constituent parts.
Schemata evolution is coarse grained relative to string evolution
because of the summing over the $N-N_2$ non-schemata bits. The evolution
of a schemata of $O(N_2)$ is described in terms of its constituent 
parts which are schematas are of order less than $N_2$. Thus the
action of crossover invokes a natural hierarchy of coarse grainings.
Such a hierarchy is reminiscent of a renormalization group 
transformation where there is a coarse graining over a subset of 
degrees of freedom, such as in the one-dimensional Ising model where
one may sum over every other spin in the partition function for 
instance. In the genetic algorithm case this coarse graining stops
naturally when one arrives at the evolution of 1-schematas as these
are not decomposable into even more coarse grained degrees of freedom.

In one sense it is remarkable that one may solve analytically a
genetic algorithm albeit for a simple fitness landscape and over a
short time interval, however, what is lacking is a reasonable
approximation scheme with which one may attack the evolution equations.
Just as solving an exact renormalization group equation is almost
impossible so with genetic algorithms finding exact solutions is probably
hopeless. However, implementing renormalization group transformations
approximately has had remarkable success in explaining many physical
phenomena. We hope that finding analogous techniques in the study
of genetic algorithms might lead to similar success.

Starting from the evolution equation for schematas, a further 
coarse-graining was performed to arrive at an expression for the 
average contribution of all schematas of size $l$ to the 
improvement of fitness. Applying this equation to the particular case 
of a $k=0$ landscape, where each bit contributes independently 
to fitness, we showed that the net effect of crossover on fitness
growth is slightly positive for all $l$: the effect of schemata 
reconstruction always exceeds that of destruction! Schematas that are either
much smaller or much larger than half the string size are most
enhanced. 

 A different situation arises if one considers a $k > 0$ landscape. In 
this case the sum of the effective selective advantages of the parts of 
a schemata is not necessarily equal to the effective selective advantage of 
the entire schemata. Only when the parts of a selected schemata are less
selected than the whole (the deceptive case), crossover leads
to a net destructive force as schematas are broken down into pieces
which are then lost due to their low selectivity. The schematas that are
selected over a long time scale are those that break down into
useful parts, independently of their size. 

 Finite size effects break the apparent symmetry of the geometrical 
effect of crossover about $l = N/2$: The existence of a selection 
threshold favours highly fit schematas with a large number of specified bits
$N_2$, and these can be found with a reasonable probability only if their
length $l$ is large. Combining this argument with the $l$-dependence of
in-schemata fitness growth $\Delta_l$ one concludes that the effective
degrees of freedom will be schematas with large $N_2 $ and $l > N/2$.

 This conclusion has important and surprising consequences for the designer of
Genetic Algorithms. It is often thought that GA designers should strive to
find a coding such that bits that ``cooperate'' are placed near each other
on the chromosome, so as to resist the destructive effect of crossover.
This is generally speaking a very difficult task, since the structure of
the optimisation problem usually does not match the linear topology of
the strings. Our results show that this task is pointless: if anything one
should try to place cooperating bits as far from each other as possible. Of 
course this is the most probable outcome if no attention is placed to 
the linear disposition of the bits, so this is not a problem one should 
worry about.

 We should stress that the above comment by no means implies that the 
choice of coding is irrelevant. The choice of a genetic interpreter is
crucial to generate a high density of states near desired fitness extrema 
and perhaps also to guide the emergence of an algorithmic language 
\citelabel{\vera}[\cite{vera}]
which facilitates the search for new highly fit schematas. These issues
however lie beyond the scope of the present paper. 

With the results of this paper in mind it is interesting to recall
the analogy between GA's and spin glass dynamics discussed in the 
introduction. In both cases one is describing a condensation process 
in a rugged landscape, guided by the emergence of overlaps with certain 
structures or ``patterns''. One of the 
chief reasons why in GA's the overlaps with schematas is considered rather 
than with entire strings ($N_2=N$ schematas) is that genetic populations 
are generally too disordered for such a rigid structure as a 
completely-specified string to be of much relevance. Of course the same
can be said of spin glasses far from equilibrium. This suggests that the 
notion of ``schemata'' may find some usage to study 
the condensation of spin glasses from an initial disordered phase. 
One can carry the analogy between GA's and spin glasses one step further 
and suggest that, in the case of sparesely-connected neural networks, the 
truncated connective trees may form a priviledged class of schematas for 
the purpose of developing an effective theory of neural dynamics.

\noindent {\bf Acknowledgements} We would like to thank Peter 
Stadler for discussions during his stay at the Institute of 
Nuclear Sciences in the Fall of 1995. This work was partially 
supported through Conacyt grant number 211085-5-0118PE.

\vfill\eject

\centerline{\bf REFERENCES}

\item{[\cite{hopfield}]} J. J. Hopfield, Proc. Natl. Sci. USA {\bf 79} (1982), 2554.

\item{[\cite{amit}]} D. Amit, H. Gutfreund and H. Sompolinsky, 
Phys. Rev. Lett. {\bf 55} (1985), 1530

\item{[\cite{zertuche}]} F. Zertuche, R. L\'opez-Pe\~na and H. Waelbroeck,
J. Phys. A {\bf 27} (1994), 5879

\item{[\cite{bak}]} P. Bak, C. Tang and K. Wiesenfeld, Phys. 
Rev. A {\bf 36} (1988), 364

\item{[\cite{genesV}]} B. Lewin, {\it Genes V\/}, Oxford University Press, Oxford (1995). 

\item{[\cite{holland}]} J. H. Holland, {\it Adaptation in natural and artificial systems\/}, MIT Press,
Cambridge, MA (1975).

\item{[\cite{goldberg}]} D. E. Goldberg, {\it Genetic Algorithms in search, 
optimization and machine learning}, Addison Wesley, Reading, MA (1989).

\item{[\cite{kaufman}]}  S. A. Kauffman, {\it The Origins of Order}, 
Oxford University Press, Oxford (1993).

\item{[\cite{mitchell}]} M. Mitchell, Complexity, 31 (1995).

\item{[\cite{taib}]}  Z. Taib, {\it Branching Processes and Neutral Evolution \/}, Springer-Verlag,
Berlin (1992). 

\item{[\cite{eigen}]}  M. Eigen, Naturwissenschaften {\bf 58}, 465 (1971); M. Eigen, 
J. McCaskill and P. Schuster, Adv. Chem. Phys. {\bf 75}, 149 (1989).

\item{[\cite{peliti}]}  L. Peliti, Lectures given at the NATO ASI on Physics of Biomaterials:
Fluctuations, Self-Assembly and Evolution, Geilo (Norway) (1995).

\item{[\cite{benn}]} A. Prugel-Bennett and J.L. Shapiro, Phys. Rev. Lett. {\bf 72}, 1305 (1994).

\item{[\cite{statan}]} I. Leuthausser, J. Chem. Phys. {\bf 84}, 1884 (1986); J. Stat. Phys. {\bf 48}, 343 (1987).

\item{[\cite{goldbridge}]}  C.L.  Bridges and D. E. Goldberg, in: 
{\it Genetic Algorithms and their Applications\/}, ed. John J. 
Grefenstette, Lawrence Erbaum Publishers (1987).

\item{[\cite{kimura}]} M. Kimura, {\it The Neutral Theory of Molecular
Evolution}, Cambridge University Press, Cambridge (1983).

\item{[\cite{vera}]}  S. Vera and H. Waelbroeck: {\it Symmetry breaking 
and adaptation: the genetic code of retroviral env proteins}. Preprint 
ICN-UNAM-96-09, adap-org/9610001 (1996).

\vfill\eject

\centerline{\bf FIGURE CAPTIONS}

\

\item{} Figure 1. The multiplicative renormalization of the effective 
fitness due to crossover, $(1 + s_l)$, is represented as a function of
the schemata length $l$. The crossover term gives a positive contribution
to fitness growth for all values of $l$, which is greater for schemata 
sizes that are either much smaller or much larger than half the chromosome
size. 

\

\item{} Figure 2. The average absolute correlations between bits in the 
chromosome are given in terms of ($B$) the linear distance which separates
the bits on the chromosome, and ($C$) the connective distance defined as 
the smallest number of landscape connections to go from one bit to the 
other. Very early on one notes a slight preference for correlations 
between bits that are near each other on the chromosome, i.e. with $l << N$
(2a). By $t=100$ the correlations between landscape-related bits become 
important (2b), and they come to dominate at $t=200$ (2c). At this point the 
population is highly organised and correlations on the basis of linear 
chromosome distance are no longer significant.

\end